%% file: Prague2013.tex
\documentclass[a4paper,twocolumn]{esapub2005} 
\usepackage{amsmath}
\usepackage{times}
\usepackage{natbib}
\usepackage{graphicx}

\newcommand{\beq}{\begin{equation}}
\newcommand{\eq}{\end{equation}}
\newcommand{\refe}[1]{\mbox{Eq.\hspace{2pt}\ref{#1}}}		
\newcommand{\reff}[1]{\mbox{Fig.\hspace{2pt}\ref{#1}}}		

\newcommand{\reft}[1]{\mbox{Table\hspace{2pt}\ref{#1}}}		
\newcommand{\refs}[1]{\mbox{Section\hspace{2pt}\ref{#1}}}		

\newcommand{\pmu}{p_{\mu}}
\newcommand{\pnu}{p_{\nu}}

\newcommand{\pfi}{p_{\varphi}}
\newcommand{\ptt}{p_t}
\newcommand{\pth}{p_{\theta}}
\newcommand{\pr}{p_r}
\newcommand{\ecc}{\ensuremath{\varepsilon}}		
\newcommand{\incl}{\ensuremath{\iota}}			

\newcommand{\pp}{\mathcal{P}}
\newcommand{\rg}{\ensuremath{r_\mathrm{g}}}

\newcommand{\parc}[2]{\frac{\partial #1}{\partial #2}}
\newcommand{\degr}{^{\circ}}

\newcommand{\QP}{\ensuremath{\hat{Q}^k(\lambda),\hat{P}_k(\lambda)}}
\newcommand{\QPi}{\ensuremath{\hat{Q}^k(\lambda_i),\hat{P}_k(\lambda_i)}}
\newcommand{\QPtau}{\ensuremath{\hat{Q}^i(\tau[k]),\hat{P}_i(\tau[k])}}
\newcommand{\QPtaubar}{\ensuremath{\hat{Q}^i(\overline{\tau}[k]),\hat{P}_i(\overline{\tau}[k])}}
\newcommand{\QPinit}{\ensuremath{\hat{Q}^i(0),\hat{P}_i(0)}}

\title{An Autonomous Reference Frame for Relativistic GNSS}
\author[1,2]{Uro\v s Kosti\' c}
\author[1]{Martin Horvat}
\affil[1]{Faculty of Mathematics and Physics, University of Ljubljana, Jadranska 19, 1000 Ljubljana, Slovenia}
\affil[2]{Slovenian centre of excellence for space sciences and technologies, A\v sker\v ceva 12, 1000 Ljubljana, Slovenia}
\author[3]{Sante Carloni}
\affil[3]{Institute of Theoretical Physics, Faculty of Mathematics and Physics, Charles University, V Holesovickach 2, 180 00 Praha 8, Czech Republic}
\author[4]{Pac\^ ome Delva}
\affil[4]{LNE-SYRTE, Observatoire de Paris, CNRS et UPMC, 61 avenue de l’Observatoire, 75014, Paris, France}
\author[1,2]{Andreja Gomboc}

\begin{document}
\maketitle
\begin{abstract}
Current GNSS systems rely on global reference frames which are fixed to the Earth (via the ground stations) so their precision and stability in time are limited by our knowledge of the Earth dynamics. These drawbacks could be avoided by giving to the constellation of satellites the possibility of constituting by itself a primary and autonomous positioning system, without any a priori realization of a terrestrial reference frame. Our work shows that it is possible to construct such a system, an Autonomous Basis of Coordinates, via emission coordinates. Here we present the idea of the Autonomous Basis of Coordinates and its implementation in the perturbed space-time of Earth, where the motion of satellites, light propagation, and gravitational perturbations are treated in the formalism of general relativity.
\end{abstract}
\section{Introduction}
\input{introduction}
\section{Motion of satellites in a perturbed space-time}
\input{hamiltonian}
%
%
\section{Relativistic positioning system}
\input{RPS}
\subsection{Autonomous Basis of Coordinates}
\input{ABC}
\section{Summary}
\input{summary}

\section*{Acknowledgments}
U.K., M.H, and A.G. acknowledge financial support from ESA PECS project \emph{Relativistic Global Navigation System}.
\bibliography{Prague2013}
\end{document}

%% file: introduction.tex
\label{sec:introduction}
The classical concept of positioning system for a Global navigation satellite system (GNSS) would work ideally if all satellites and the receiver were at rest in an inertial reference frame. But at the level of precision needed by a GNSS, one has to consider curvature and relativistic inertial effects of spacetime, which are far from being negligible. These effects are most easily and elegantly dealt with in a relativistic positioning system based on emission coordinates \citep{coll91,rovelli02,blagojevic02,vCadevz2010,Delva2011}. They depend on the set of four satellites and their dynamics, and can be linked to a terrestrial reference system. Consequently, the difficulty no longer lies in the conception of the primary reference frame but in its link with terrestrial reference frames \citep{vCadevz2010}. This allows to control much more precisely all the perturbations that limit the accuracy and the stability of the primary reference frame, if the dynamics of the GNSS satellites, described by their orbital parameters, is known sufficiently well.

Our previous work shows that it is possible to construct such a system and do the positioning within it: the orbital parameters of the GNSS satellites can be determined and checked internally by the GNSS system itself through inter-satellite links \citep{vCadevz2011}. In this way, we can construct an Autonomous Basis of Coordinates (ABC) which is independent of any Earth based coordinate system. This system constructs itself as each satellite receives proper times of all other satellites, and by comparison with its own proper time determines the parameters of the ABC system with great accuracy.

Here we present the results of our recent work, where we have further developed relativistic positioning and the ABC by including all relevant gravitational perturbations, such as Earth multipoles (up to the 6th), Earth solid and ocean tides, the Sun, the Moon, Jupiter, Venus, and the Kerr effect.

%% file: hamiltonian.tex
\label{sec:hamiltonian}
The perturbed satellite orbits were calculated from the perturbed Hamiltonian \citep{Goldstein1980}
\beq
H=\frac{1}{2} g^{\mu\nu} p_\mu p_\nu\ ,
\label{eq:hamiltonian0}
\eq
where the perturbed metric $g_{\mu\nu}$ is a sum of the unperturbed (i.e. Schwarzschild) metric $g_{\mu\nu}^{(0)}$ and the perturbative metric $h_{\mu\nu}$
\beq
g_{\mu\nu} = g_{\mu\nu}^{(0)} + h_{\mu\nu}\ ,\hspace{0.5cm}|h_{\mu\nu}|\ll g_{\mu\nu}^{(0)}\ .
\eq
The Hamiltonian (\refe{eq:hamiltonian0}) is rewritten with the canonical momenta $p_{\mu}=g_{\mu\nu}\dot x^\nu$ as
\beq
H=\frac{1}{2} g^{(0)\mu\nu} p_\mu p_\nu - \frac{1}{2} h^{\mu\nu} p_\mu p_\nu =
H^{(0)} + \Delta H\ ,
\eq
where $H^{(0)}$ and $\Delta H$ are the unperturbed and the perturbative part of the Hamiltonian, respectively.

The unperturbed orbit is then given by the equations
\beq
 \dot x^\mu  = \frac{\partial H^{(0)}}{\partial p_\mu^{(0)}} \hspace{0.5cm} \mathrm{and}\hspace{0.5cm}
  \dot p_\mu^{(0)}  = -\frac{\partial H^{(0)}}{\partial x^\mu} \ ,
\eq
where the coordinates and the momenta for Schwarzschild case are $(x^{\mu},\pmu)=(t,r,\theta,\phi,\ptt,\pr,\pth,\pfi)$. The unperturbed Hamiltonian $H^{(0)}$ admits 6 constants of motion $(Q^k,P_k)$: the orbital energy $P_1 = E$, the magnitude of the angular momentum $P_2 = l$, the $z$-component of the angular momentum $P_3 = l_z$, the time of the first apoapsis passage $Q^1 = -t_a$, the longitude of the apoapsis $Q^2= \omega$, and the longitude of the ascending node $Q^3=\Omega$. These constants are used to find the analytical solutions for the orbits in the form:
\beq
\begin{split}
t &=t(\lambda|Q^k,P_k)\hspace{0.5cm}
r =r(\lambda|Q^k,P_k)\\
\theta &=\theta(\lambda|Q^k,P_k)\hspace{0.5cm}
\phi =\phi(\lambda|Q^k,P_k)\ ,
\end{split}
\label{eq:unpert_solutions}
\eq
where $\lambda$ is the true anomaly \citep{Kostic2012}.

If the metric is perturbed, $(Q^k,P_k)$ are no longer constants of motion -- they are slowly changing functions of time. Labelling these new canonical variables as $(\hat{Q}^k,\hat{P}_k)$, we obtain their time evolution from the following expressions \citep{Goldstein1980}:
\begin{align}
\hat{\dot{Q}}^k &\simeq \left . \parc{\Delta H}{P_k}\right\vert_{Q^k,P_k}  = -\frac{1}{2}\parc{h^{\mu\nu}\pmu\pnu}{P_k}\label{eq:derivatives1}\\
\hat{\dot{P}}_k &\simeq - \left .\parc{\Delta H}{Q^k}\right\vert_{Q^k,P_k} = \frac{1}{2}\parc{h^{\mu\nu}\pmu\pnu}{Q^k}\ .\label{eq:derivatives2}
\end{align}
The \mbox{Eqs.\hspace{2pt}\ref{eq:derivatives1}} and \ref{eq:derivatives2} are numerically integrated to obtain the solutions for $\hat{Q}^k(\lambda|Q^k,P_k)$ and $\hat{P}_k(\lambda|Q^k,P_k)$, which are then used to replace $(Q^k,P_k)$ in the analytical expressions for unperturbed orbits (\refe{eq:unpert_solutions}). In this way, the perturbed orbit is described as time-evolving unperturbed orbit:
\beq
\begin{split}
t &=t(\lambda|\QP)\hspace{0.5cm}
r =r(\lambda|\QP)\\
\theta &=\theta(\lambda|\QP)\hspace{0.5cm}
\phi =\phi(\lambda|\QP)\ .
\end{split}
\label{eq:pert_solutions}
\eq

Before the \mbox{Eqs.\hspace{2pt}\ref{eq:derivatives1}} and \ref{eq:derivatives2} are solved numerically, we calculate the metric $h_{\mu\nu}$ for Kerr effect, Jupiter, Venus, the Moon, the Sun, Earth multipoles, solid, and ocean tides using relativistic multipole expansion \citep{Brdo,paper1}.

\begin{figure}
\includegraphics[width=\linewidth]{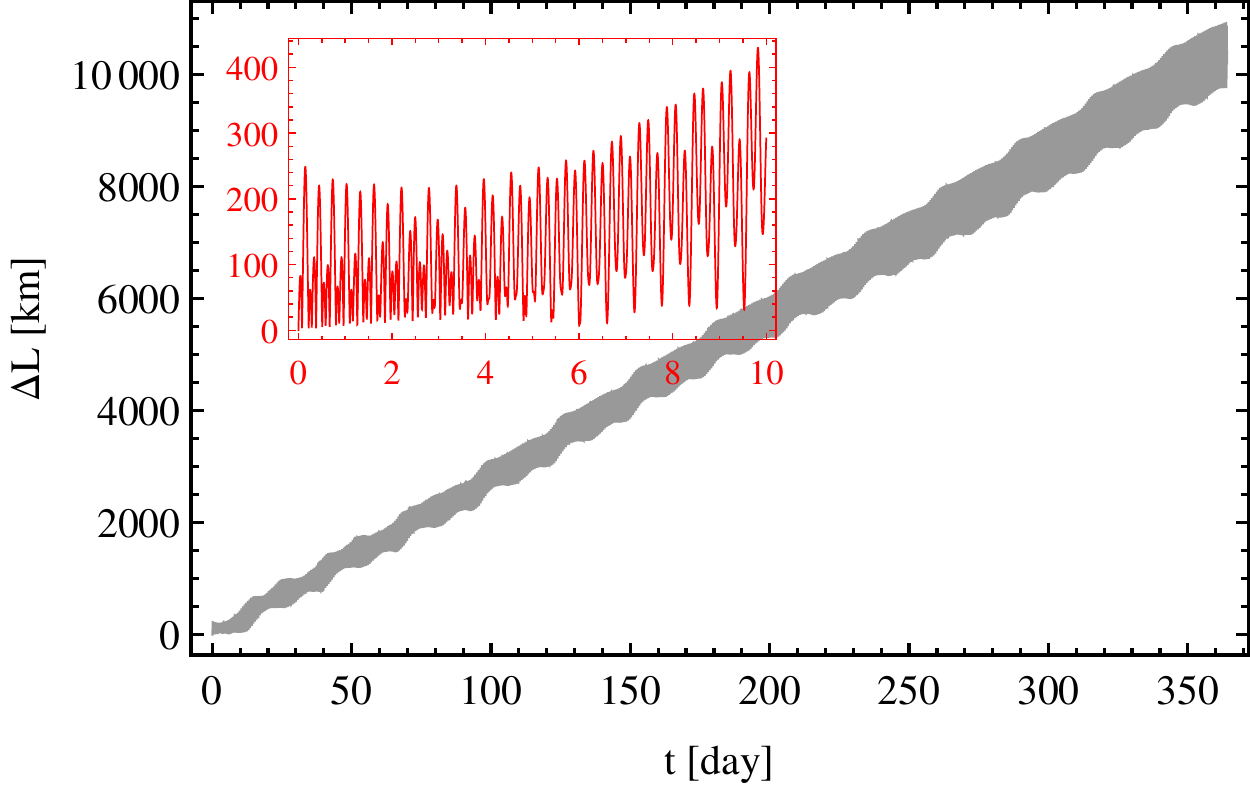}
\caption{The differences in the position $\Delta L$ of the satellite due to sum of all perturbations. The gray graph shows the long-term changes in one year, while the red inset shows the short time-scale changes within the first 10 days. The (Schwarzschild) time on $x$-axis counts days from 1 January 2012 at 12:00 noon. Axes on red and gray plots have the same units. The initial values of parameters are: $t_a = 12$ h, $\omega=0^{\circ}$, $\Omega = 0^{\circ}$, $a=29602$ km, $\ecc = 0.007$, $\incl=56^{\circ}$, corresponding to the Galileo system.}
\label{fig:everything}
\end{figure}

We calculated the evolution of orbital parameters (with initial values $t_a = 12$ h, $\omega=0^{\circ}$, $\Omega = 0^{\circ}$, $a=29602$ km, $\ecc = 0.007$, $\incl=56^{\circ}$, corresponding to the Galileo system) for each perturbation individually as well as for the sum of all perturbations and thus obtained the corresponding orbits. Because the perturbed orbits do not differ from the unperturbed ones significantly, in \reff{fig:everything} we show only the differences between the perturbed and unperturbed positions for all perturbations included: the difference in positions $\Delta L$ are $\sim 300$~km in 10 days and $\sim 10000$ km in one year. A detailed study of effects of each perturbation on each orbital parameter shows that these changes arise mostly from the precessions of $\Omega$ and $\omega$; the changes due to other orbital parameters are 2-3 orders of magnitude smaller and oscillating.
%
%
%

%% file: RPS.tex
\label{sec:RPS}
We simulate a constellation of four satellites moving along their time-like geodesics \citep{vCadevz2010,Delva2011,Kostic2012}. The initial orbital parameters of the geodesics are known and their evolution due to perturbations is calculated as shown in \refs{sec:hamiltonian}.

At every time-step of the simulation, each satellite emits a signal and a user on Earth receives signals from all satellites -- the signals are the proper times of satellites at their emission events and constitute the emission coordinates of the user. The emission coordinates determine the user's ``position'' in this particular relativistic reference frame defined by the four satellites and allow him to calculate his position and time in the more customary Schwarzschild coordinates. So, to simulate the positioning system, two main algorithms have to be implemented: (1) determination of the emission coordinates, and (2) calculation of the Schwarzschild coordinates.
\paragraph{Determination of the emission coordinates} The satellites' trajectories are parametrized by their true anomaly $\lambda$. The event $\pp_o = (t_o,x_o,y_o,z_o)$ marks user's Schwarzschild coordinates at the moment of reception of the signals from four satellites. Each satellite emitted a signal at event $\pp_i = (t_i,x_i,y_i,z_i)$, corresponding to $\lambda_i$ ($i=1,...,4$). Emission coordinates of the user at $\pp_o$ are, therefore, the proper times $\tau_i (\lambda_i)$ of the satellites at $\pp_i$. Taking into account that the events $\pp_o$ and $\pp_i$ are connected with a light-like geodesic,\footnote{The light-like geodesics are calculated in Schwarzschild space-time without perturbations, because the effects of perturbations on light propagation are negligible.}  we calculate $\lambda_i$ at the emission point $\pp_i$ using the equation
\beq
\begin{split}
t_o - t_i(\lambda_i|\QPi) &=\\
T_f(\vec{R}_i(\lambda_i|&\QPi), \vec{R}_o)\ ,
\label{eq:positioning}
\end{split}
\eq
where $\vec{R}_i=(x_i,y_i,z_i)$ and $\vec{R}_o = (x_o,y_o,z_o)$ are the spatial vectors of the satellites and the user, respectively. The function $T_f$ calculates the time-of-flight of photons between $\pp_o$ and $\pp_i$ as shown by \citet{2005PhRvD..72j4024C} and \citet{vCadevz2010}. The \refe{eq:positioning} is actually a system of four equations for four unknown $\lambda_i$ -- once the values of $\lambda_i$ are determined, it is straightforward to calculate $\tau_i$ for each satellite and thus obtain user's emission coordinates at $\pp_o = (\tau_1,\tau_2,\tau_3,\tau_4)$.
\paragraph{Calculation of the Schwarzschild coordinates} Here we solve the inverse problem of calculating Schwarzschild coordinates of the event $\pp_o$ from proper times $(\tau_1,\tau_2,\tau_3,\tau_4)$ sent by the four satellites. We do this in the following way: For each satellite, we numerically solve the equation
\beq
\tau(\lambda_i|\QPi) = \tau_i\ ,
\eq
to obtain $\lambda_i$, where $\tau(\lambda|\QP)$ is a known function for proper time on time-like geodesics \citep{Kostic2012}. The Schwarzschild coordinates of the satellites are then calculated from $\lambda_i$ using \refe{eq:pert_solutions}. With the satellites' coordinates known, we can take the geometrical approach presented by \citet{vCadevz2010} to calculate the Schwarzschild coordinates of the user. The final step in this method requires us again to solve \refe{eq:positioning}, however, this time it is treated as a system of 4 equations for 4 unknown user coordinates, i.e., solving it, gives $(t_o,x_o,y_o,z_o)$.

The accuracy of this algorithm has been tested for satellites on orbits with initial parameters given in \reft{tab:parameters} and a user at coordinates $r_o=6371$~km, $\theta_o=43.97\degr$, $\phi_o=14.5\degr$. The user's coordinates remain constant during the simulation. The relative errors, defined as
\beq
\epsilon_t  = \frac{ t_o - t_o^e}{t_o}, \
\epsilon_{x,y,z} = \frac{ \vec{R}_o - \vec{R}_o^e}{\vec{R}_o}\ ,
\eq
are of the order $10^{-32}-10^{-30}$ for coordinate $t$, and $10^{-28}-10^{-26}$ for $x$, $y$, and $z$; here $t_o^e$ and $\vec{R}_o^e$ are user time and coordinates as calculated from the emission coordinates. Using a laptop\footnote{With the following configuration: Intel(R) Core(TM) i7-3610QM CPU @ 2.30GHz, 8GB RAM, Intel C/C++/Fortran compiler 13.0.1.} for calculations, the user's position (with such errors) was determined in 0.04 s, where we assumed that (1) in real applications of the positioning the true values of orbital parameters would be transmitted to the user together with the emission coordinates, so to account for this in our simulations, we calculated the evolution of parameters from their initial values before starting the positioning, and (2) the position of the user is completely unknown, i.e., we do not start from the last known position. If we did, the times for calculating the position would be even shorter.
\begin{table}
  \centering
    \caption{Orbital parameters for 4 satellites: longitude of ascending node ($\Omega$), longitude of perigee ($\omega$), inclination (\incl), major semi-axis ($a$), eccentricity (\ecc ), and time of apogee passage ($t_a$).}\vspace{1em}
    \renewcommand{\arraystretch}{1.2}
    \begin{tabular}[h]{ccccccc}
      \hline
      \# & $\Omega\ [ \degr]$ & $\omega\ [ \degr]$ & \incl $\ [ \degr]$ & $a\ [\mathrm{km}]$ & \ecc & $t_a\ [\mathrm{s}]$\\
      \hline
		1 & 0 & 270 & 45 & 30000 & 0.007 & 0\\
		2 & 0 & 315 & 45 & 30000 & 0.007 & 0\\
		3 & 0 & 275 & 135 & 30000 & 0.007 & 0\\
		4 & 0 & 320 & 135 & 30000 & 0.007 & 0\\
      \hline
      \end{tabular}
    \label{tab:parameters}
\end{table}

%% file: ABC.tex
\label{sec:ABC}
To construct an autonomous coordinate system, we apply the idea of the Autonomous Basis of Coordinates (ABC) presented in \citet{vCadevz2011} to a perturbed satellite system, i.e., we simulate the motion of a pair of satellites along their perturbed orbits.

\begin{figure}
\centering
\includegraphics[width=0.9\linewidth]{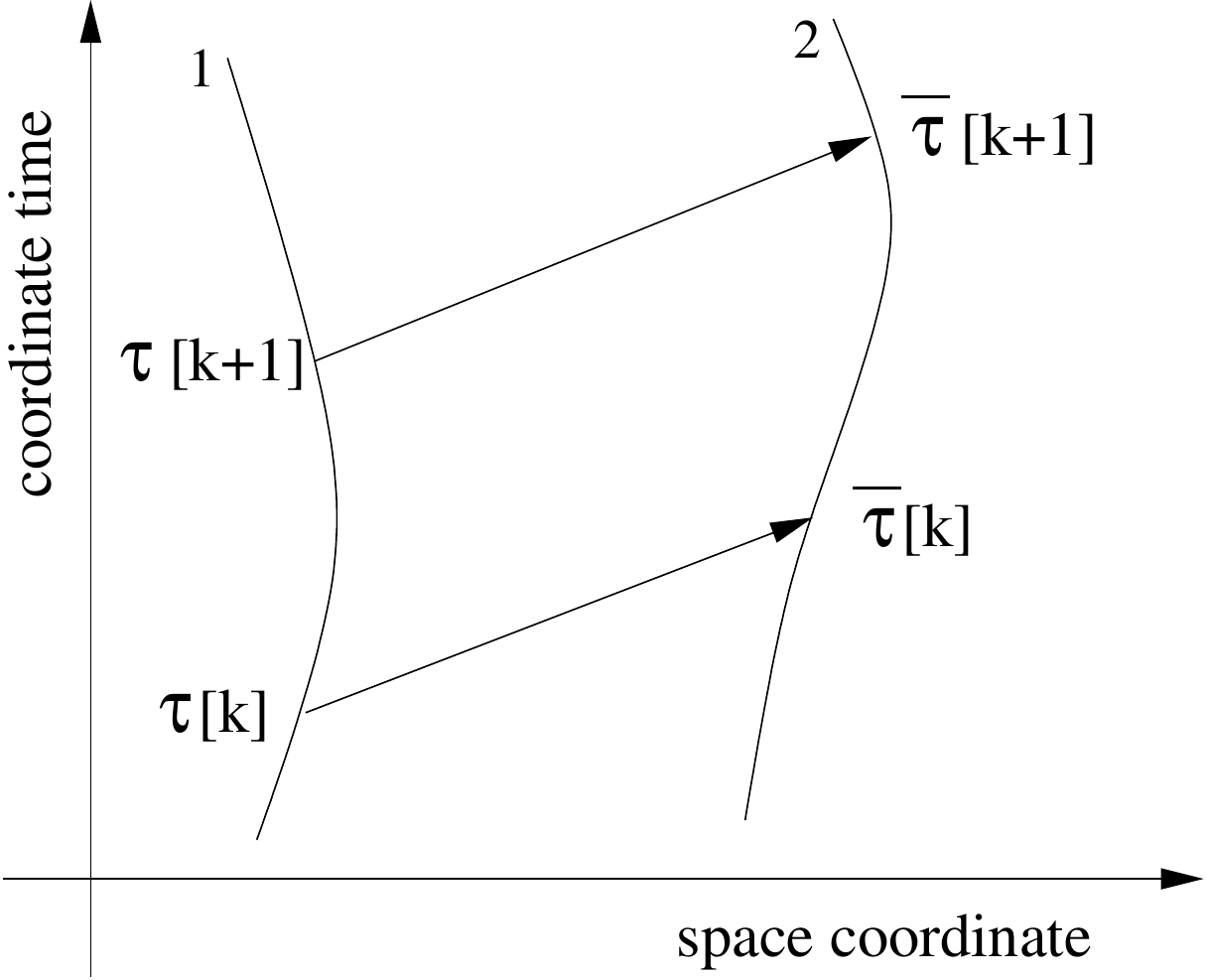}
\caption{A pair of satellites exchanging their proper times. At every time-step $k$, the satellite 1 sends the proper time of emission $\tau[k]$ to the satellite 2, which receives it at the time of reception $\overline{\tau}[k]$. The emission and reception event pairs are connected with a light-like geodesic.}
\label{fig:pairs}
\end{figure}

At each time-step of the simulation, both satellites exchange emission coordinates as shown in \reff{fig:pairs}, where, for clarity, only communication from satellite 1 to satellite 2 is plotted. 
These events of emission at proper time $\tau$ of the first satellite and reception at $\overline{\tau}$ of the second satellite are connected with a light-like geodesic, i.e., the difference between the coordinate times of emission $t_1(\tau)$ and reception $t_2(\overline{\tau})$ must be equal to the time of flight of a photon between the two satellites (cf. \refe{eq:positioning})
\beq
T_f=t_2(\overline{\tau}) - t_1(\tau)\ .
\label{eq:differences}
\eq
However, this is only true if we know the exact values of the initial orbital parameters of each satellite, as well as their evolution.

When constructing the relativistic positioning system, it is reasonable to assume that the initial orbital parameters $(\QPinit)$ are not known very precisely. To improve their values, we sum the differences between RHS and LHS of \refe{eq:differences} for all communication events into an action
\beq
\begin{split}
S(\QPinit) = &\sum_k \left(t_1(\tau[k]|\QPtau) -\right.\\
 &t_2(\overline{\tau}[k]|\QPtaubar) - \\
			&T_f(\vec{R}_1(\tau[k]|\QPtau),\\
			&\left. \vec{R}_2(\overline{\tau}[k]|\QPtaubar))\right)^2\ ,
\end{split}
\eq
which has a minimum value (close to zero) for the true initial values of orbital parameters -- for the $2\times 6$ orbital parameters that we have ($\QPinit$ for both satellites), this becomes a problem of finding a minimum of a 12D function. Because the orbital parameters depend on time, their time evolution has to be recalculated (as presented in \refs{sec:hamiltonian}) at every step of the minimization, which makes the minimization process very slow.

\begin{figure}
\centering
\includegraphics[width=\linewidth]{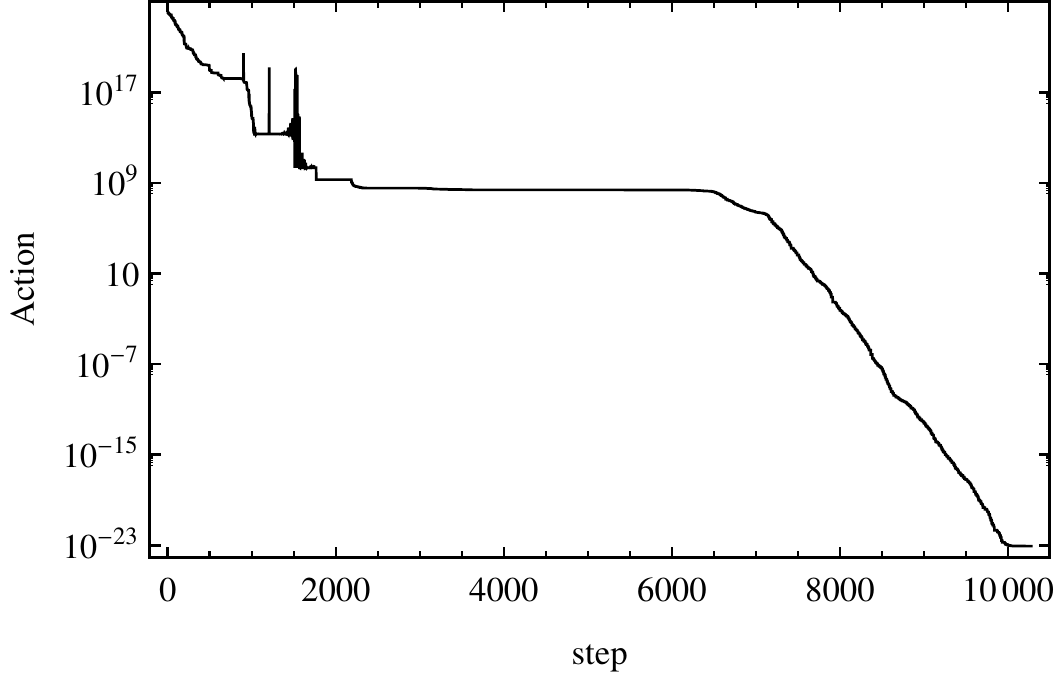}
\caption{The action $S(\QPinit)$ during the minimization process. The first stage takes 1765 steps and the second one takes 8513 steps.}
\label{fig:minimization}
\end{figure}

The minimization was done in two stages. In the first stage, we use the PRAXIS minimization method \citep{brent1973algorithms} implemented in the NLOPT library \citep{NLOPT} to determine the parameters within double precision. The resulting values are then used as initial values for the second stage, where we use the simplex method to ``polish'' the parameters within 128-bit quad precision.\footnote{Quad precision is required if the resulting parameters are used in \refe{eq:pert_solutions}, where cancellation effects become significant in case of quasi-circular orbits. 
} In \reff{fig:minimization} we plot the values of the action during the minimization process; the first stage takes 1765 steps, while the second one takes 8513 steps.

The number of time-steps along the orbits was sufficiently large ($k=1\ldots 433$) to cover approximately two orbital times. The initial values of the orbital parameters used as starting point in the minimization differ from the true values by an amount which induces the error of $\sim 2-3$~km in the satellites' positions. At the beginning of the minimization, the value of the action is $S \approx 10^{24}(\rg/c)^2$, at the end of the first stage it is $2\times 10^{10}(\rg/c)^2$, and at the end of the second stage it drops to $8\times 10^{-24}(\rg/c)^2$. The relative errors of the orbital parameters $(\QPinit)$ after the minimization are of the order of $10^{-22}$.

By repeating the minimization procedure for all possible pairs of satellites, we can reconstruct the orbital parameters of every satellite in the system \emph{without tracking the satellites from the Earth} and thus obtain an autonomous coordinate system.

%% file: summary.tex
\label{sec:summary}
In this contribution we have shown how to construct an Autonomous Basis of Coordinates (ABC) for relativistic GNSS in a perturbed Schwarzschild space-time.

The ABC concept establishes a local inertial frame, which is based solely on dynamics of GNSS satellites and is thus completely independent of a terrestrial reference. General relativity was used to calculate the dynamics of the GNSS satellites and the gravitational perturbations affecting the dynamics (Earth multipoles, Earth solid and ocean tides, the Sun, the Moon, Jupiter, Venus, and the Kerr effect), as well as the inter-satellite communication, which actually makes possible for construction of the whole system. The numerical codes for positioning determine all four coordinates in 40~ms with 25-30 digit accuracy, showing that general relativistic treatment presents no technical obstacles for current GPS devices.

Because the system does not rely on Earth based reference frames and is constructed only through proper time exchange between satellites, it offers unprecedented accuracy and stability. In fact, present technology, planned to be used in the Galileo system, should be able to routinely reach millimetre accuracy with respect to an absolute local inertial frame defined independently of Earth based coordinates. At this level of accuracy it seems necessary to decouple the local inertial frame from the geodetic Earth frame, and allow the comparison of the two, to tell us fine details about Earth rotation, gravitational potential and dynamics of Earth crust. Last but not least, tracking of satellites with ground stations is necessary only to link the relativistic positioning system to a terrestrial frame, although this link can also be obtained by placing several receivers at the known terrestrial positions.